\documentclass{vgtc}                          %

\ifpdf%
  \pdfoutput=1\relax                   %
  \usepackage{graphicx}                %
  \DeclareGraphicsExtensions{.pdf,.png,.jpg,.jpeg} %
\else%
  \usepackage{graphicx}                %
  \DeclareGraphicsExtensions{.eps}     %
\fi%

\graphicspath{{figures/}{pictures/}{images/}{./}} %

\usepackage{microtype}                 %
\PassOptionsToPackage{warn}{textcomp}  %
\usepackage{textcomp}                  %
\usepackage{mathptmx}                  %
\usepackage{times}                     %
\usepackage{cite}                      %
\usepackage{tabu}                      %
\usepackage{booktabs}                  %
\usepackage{float}
\usepackage{xcolor}
 \usepackage{enumitem}

\usepackage{comment}

\usepackage{epsfig}
\usepackage{amssymb}
\usepackage{amsmath}
\usepackage{amsfonts}
 \usepackage{enumitem}
\usepackage{balance}

\def\widthnarrow{241.14749pt}

\newcommand{\fone}[2]{$F_{1,\textnormal{#1}}=#2$}

\onlineid{0}

\vgtccategory{Research}

\vgtcinsertpkg

\def\projname{TapID}

\def\tapidconfig{\textsc{TapID}}
\def\fitbitconfig{\textsc{Fitbit}}
\def\smartwatchconfig{\textsc{Smartwatch}}
\def\allfourconfig{\textsc{All~4}}
\def\timestamp{timestamp}

\title{\projname: Rapid Touch Interaction in Virtual Reality using Wearable Sensing}

\author{Manuel Meier\thanks{e-mail addresses: firstname.lastname@inf.ethz.ch} %
\and Paul Streli$^*$ %
\and Andreas Fender$^*$ %
\and Christian Holz$^*$
}
\affiliation{\scriptsize Department of Computer Science \\ ETH Zürich, Switzerland}

\def\appphoto{Photo sorting}
\def\appdocedit{Document editing}
\def\appnumberinput{Number input}
\def\apppiano{Piano}

\newcommand\FIGApps[1]{
    \begin{figure*}[#1]
    \includegraphics[width=\textwidth]{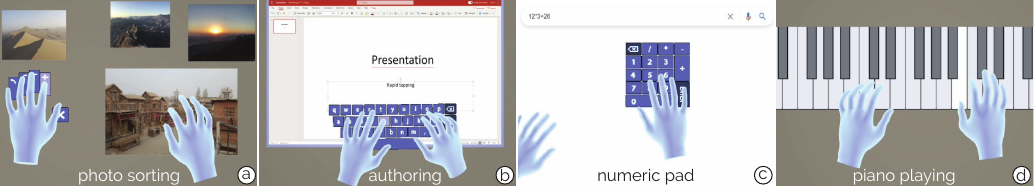}
    \caption{%
        Our demo applications showcase \projname's integration with VR scenes and hand-based interaction using different surface widgets:
        (a)~dragging and resizing photos including a per-finger command input widget below the left hand,
        (b)~editing rich-text documents with keyboard input including text cursor navigation,
        (c)~a numeric keypad for inputting numbers and operations, and
        (d)~virtual piano keys for precise input.
    }
    \label{fig:applications}
    \end{figure*}
}

\def\teaserA{a}
\def\teaserB{b}
\def\teaserC{c}
\def\teaserD{d}

\newcommand\teaserref[1]{Figure~\ref{fig:teaser}#1}

\def\teaserrefA{\teaserref{\teaserA}}
\def\teaserrefB{\teaserref{\teaserB}}
\def\teaserrefC{\teaserref{\teaserC}}
\def\teaserrefD{\teaserref{\teaserD}}

\teaser{
    \centering
    \includegraphics[width=\textwidth]{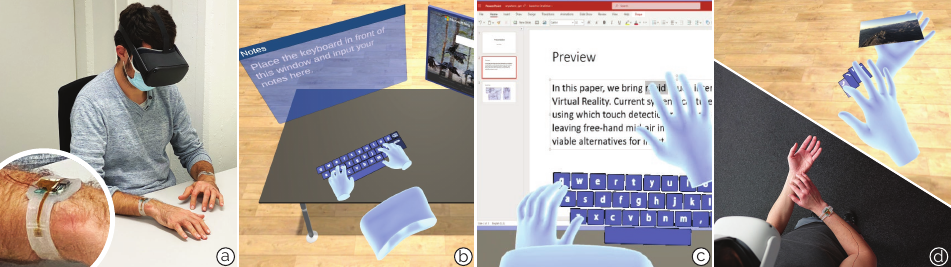}
        \caption{
        \projname{} is a wrist-worn device that detects taps on surfaces and identifies the tapping finger, which, combined with tracked hand poses, triggers input in VR.
        (a)~The user is wearing two \projname{} bands for
        (b)~touch interaction with surface widgets in VR, e.g., for text input, web browsing, or (c)~document authoring using familiar front-end apps.
        (d)~Widgets can also be registered to the body itself, using \projname\ to detect on-body taps and identify the tapping finger, here to rotate an image held in hand.
        }
    \label{fig:teaser}
}

\abstract{
Current Virtual Reality systems typically use cameras to capture user input from controllers or free-hand mid-air interaction.
In this paper, we argue that this is a key impediment to productivity scenarios in VR, which require continued interaction over \textit{prolonged} periods of time---a requirement that controller or free-hand input in mid-air does not satisfy.
To address this challenge, we bring rapid touch interaction on surfaces to Virtual Reality---the input modality that users have grown used to on phones and tablets for continued use.
We present TapID, a wrist-based inertial sensing system that complements headset-tracked hand poses to trigger input in VR.
TapID embeds a pair of inertial sensors in a flexible strap, one at either side of the wrist; from the combination of registered signals, TapID reliably detects surface touch events and, more importantly, \textit{identifies} the finger used for touch.
We evaluated TapID in a series of user studies on event-detection accuracy ($F_1=0.997$) and hand-agnostic finger-identification accuracy (within-user: $F_1=0.93$; across users: $F_1=0.91$ after $10$ refinement taps and $F_1=0.87$ without refinement) in a seated table scenario.
We conclude with a series of applications that complement hand tracking with touch input and that are uniquely enabled by TapID, including UI control, rapid keyboard typing and piano playing, as well as surface gestures.%
}

\CCScatlist{
  \CCScatTwelve{Human-centered computing}{Human computer interaction (HCI)}{Interaction paradigms}{Virtual reality};
  \CCScatTwelve{Human-centered computing}{Human computer interaction (HCI)}{Interaction techniques}{Gestural input}
}

\usepackage[disable]{todonotes}

\begin{document}

\maketitle

\section{Introduction}
The latest Mixed Reality systems incorporate hand pose recognition for input detection, for Augmented Reality (e.g., Hololens 2) and Virtual Reality (e.g., Quest 2) alike.
While previous device generations had largely relied on hand-held controllers for input, the transition to controller-free, hand pose and gesture-operated input is apparent, both in the research communities (e.g., \cite{xiao2018mrtouch}) as well as the commercial domain (e.g., \cite{hololens2,quest2,vive}).
 
In Augmented Reality, hand tracking is well-suited to accompany the work on physical objects, for example in maintenance and repair~\cite{henderson2010exploring} or manufacturing and assembly~\cite{tang2003comparative}.
Working on physical objects allows the user to hold onto something, thus preventing fatigue during use~\cite{hincapie2014consumed}.
In contrast, for the interaction with objects in Virtual Reality, where content is intangible outside passive haptics systems (e.g., \cite{insko2001passive,kohli2010redirected,cheng2017sparse}), free-hand interaction can become a burden once it exceeds quick interactions~\cite{jang2017modeling}.

In this paper, we propose moving all \textit{direct} interaction in VR to passive surfaces.
Compared to mid-air interaction, touch interaction on surfaces provides users with an opportunity to rest their arms between interactions while  simultaneously offering physical support during prolonged interactions.
At the same time, the surface provides haptic feedback alongside a frame of reference for proprioception, affording quick and precise interaction.

The core challenge of touch interaction in VR systems is the precise detection of touch events; because hands are tracked from the head-mounted systems themselves, this single vantage point can lead to inaccuracy in depth sensing and thus make touch/no-touch discrimination and contact locating~\cite{agarwal2007high} challenging.
A variety of approaches have been proposed to detect touch on surfaces using depth cameras, for example for stationary~\cite{wilson2010using} and headset-integrated systems~\cite{xiao2018mrtouch}.
To combat noise in-depth measurements, they integrate filters for reliable touch detection and rejection.
Both, adding filters as well as the framerate of the used camera inherently limit the speed of interaction that systems can reliably detect.

We address the challenge of reliable and quick touch detection with our band \projname{}, which users wear around their wrist.
Through built-in inertial sensors, \projname{} does not just detect touch events on passive surfaces but additionally identifies the finger used for touch.
\projname{} thus complements the headset's optical tracking of hands and fingers, with which we combine \projname's detected touches to trigger input events in VR.
Our approach, therefore, allows applications and interaction modalities known from tablets and phones to seamlessly and reliably transfer to VR scenarios.

\begin{figure}[h]
    \centering
    \includegraphics[width=\widthnarrow]{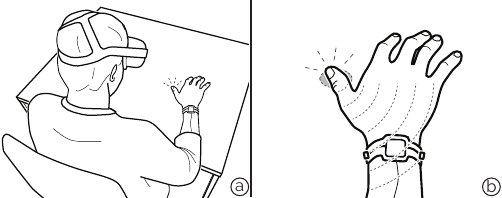}
    \caption{\projname's sensing principle.
    The user interacts through touches on the table.
    (a)~\projname\ detects the occurrence of a tap and identifies the tapping finger, while the VR headset tracks fingertip positions. %
    (b)~A tap is mechanically propagating through the hand and is registered by \projname's dual-IMU sensor, here identifying the thumb.}
    \label{fig:device_preview}
\end{figure}

\subsection{Rapid touch interaction on passive surfaces in VR}

Figure~\ref{fig:teaser} shows an overview of our approach.
Here, a user is interacting through freehand gestures in VR, while \projname{} supplies the tracking signals for finger touches on the table in front of them.
The user thereby wears one \projname{} band on either wrist, which streams inertial events to a PC for tap detection and finger identification.

Simultaneously, the VR headset tracks the wearer's hands using built-in vendor algorithms that process the headset's camera feeds.
Our system then obtains the 3D-tracked hand poses from the headset for both hands when they are in the view frustrum. %

When \projname\ detects tap events through the inertial sensors, our system forwards the input event to the VR application.
After verifying whether the user's hands are within proximity of a passive surface such as a table, our VR app then triggers an input event at the corresponding finger's location.
If the user's hands are not close to a surface, our system simply rejects the event detected by TapID as inadvertent input and ignores it.

\teaserrefB\ shows an application that particularly benefits from our approach: a keyboard that affords rapid and reliable text entry using all ten fingers.
Upon typing on the keyboard, our system forwards all key input to the operating system for text processing and retrieves word suggestions that the user may select---similar to how they would interact on a regular touch tablet.
Our system integrates a browser interface and forwards all input to the renderer, so that any web application such as search engines, word processors, or presentation apps can become usable through touch in VR using \projname.
As shown in \teaserrefC, Microsoft PowerPoint is running in VR and supports all the functionality users would expect from touch interfaces, such as tapping to place the cursor, double-tapping to select a word, tool buttons, etc.
Finally, \teaserrefD\ shows an example application that uses the user's own body for input affordances, for example when passive surfaces around the user are absent.
Here, menus and other interactive elements are registered with the user's arm, tracked by the headset, and a tap on the body delivers input to the UI elements, as it is picked up and processed by \projname.

\begin{figure}[t]
    \centering
    \includegraphics[width=\columnwidth]{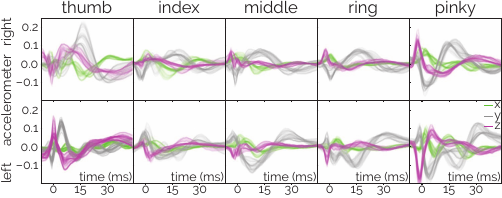}
        \caption{Overlaid inertial measurements from the IMUs on the left and right side of the wrist, respectively, for 30 sample taps caused by each of a user's 5 fingers of the right hand.}
    \label{fig:event_detection}
\end{figure}

Figure~\ref{fig:device_preview} illustrates our sensing principle. 
The elastic silicone-based strap embeds our SoC platform that connects to two accelerometers, one on either side of the wrist. %
The accelerometers mechanically couple to the wearer's skin and wrist bones through the strap and, thus, pick up body-coupled vibration signals.
During each tap event, the inertial sensors register slightly different vibration signals due to the propagation path through the user's hand depending on the finger they originate from.
These differences are subtle yet systematic as shown in Figure~\ref{fig:event_detection}, and they are the source of our signal, which \projname{} analyzes using a machine-learning pipeline to detect taps and identify the tapping finger.

In our user study, we evaluated \projname's performance of tap event detection and finger identification with 18 participants.
\projname{} reliably detected tap events (\fone{tap}{.996}) and reliably identified the tapping finger (cross-session \fone{finger}{.93}, cross-person with refinement \fone{finger}{.91}, cross-person without refinement \fone{finger}{.87}).

We compared our results to the performance of our method running on a rigid and \textit{board-mounted} dual-accelerometer configuration (e.g., found in \fitbitconfig{} bands, such as Charge~\cite{fitbit_charge}), where it achieved comparable accuracies for tap detection \fone{tap}{.996} and worse but still useful finger identification (cross-session \fone{finger}{.92}, cross-person with refinement \fone{finger}{.89}, without refinement \fone{finger}{.80}).
We also evaluated our prototype in a traditional \smartwatchconfig{} configuration with only a single board-mounted accelerometer. This setup achieved comparable results to the \fitbitconfig{} for both tap detection (\fone{tap}{.996}) and finger identification (cross-session \fone{finger}{.91}, cross-person with refinement \fone{finger}{.89}, without refinement \fone{finger}{.81}).
These promising results, particularly in the configuration of existing wearable devices, validate our method for the practical use in VR scenarios.

\subsection{Contributions}

We make three main contributions with our work.
1)~We introduce a wrist-based sensing principle to detect surface touches and to identify the finger that has caused a touch through a dual-accelerometer configuration integrated into a watch strap.
2)~We demonstrate our prototype implementation \projname\ that processes and classifies input events in real-time using a machine-learning pipeline.
The detected and classified events are then forwarded to our VR system that combines them with the hand posed tracked in 3D by the headset to trigger input in VR.
3)~In a technical evaluation with 18 participants, we assess \projname's performance in detecting tap events and correctly identifying fingers.
We thereby compare \projname{} in three configurations: the proposed dual-IMU implementation on the wrist strap, a dual-IMU sensor rigidly mounted on the board (\fitbitconfig), and a single on-board accelerometer (\smartwatchconfig), investigating accuracies cross-session and cross-person.
Notably, we demonstrate that our method is \emph{neither person-specific nor hand-specific}.

Collectively, these contributions highlight the promise of touch interaction in VR, using the input modality that is familiar from tablet and phone devices, but extending and integrating it into the spatial user interfaces common in VR.
We believe that our method has the potential to bring prolonged interaction to VR in the context of creativity and productivity scenarios, where reliable, precise, and non-fatiguing input takes precedence over the shorter and often whole-body engaging interactions that have already proven successful in the context of games and entertainment.

\newcommand\rwsection[1]{%
\subsection{#1}%
}%

\section{Related Work}

\projname{} is related to efforts on input event detection, particularly using wearable sensors, finger identification, and VR interaction.

\rwsection{Wearable sensors and body-coupled events}
Several projects have investigated body-coupled mechanical events for interaction.
Approaches mainly differ in the origin of such events, varying between events that result from hand-held tools, repetitive motions during an activity, contact with passive affordances, and mechanical events caused by tapping one's own body.

Inertial measurement units (IMUs) are frequently used to detect human motion and activity due to their integration into phones and wearable devices~\cite{wang2019deep}. 
For example, Viband operates the IMU in a watch at a high sample rate to detect handheld tools and manual activities~\cite{laput2016viband}.
With the emergence of neural network-based processing pipelines, various network architectures have been presented for such
recognition~\cite{zeng2014convolutional,murad2017deep,ordonez2016deep,jiang2015human},
which has started to supersede the previously hand-crafted features (e.g.,~\cite{janidarmian2017comprehensive,espinilla2018human}).

To sense explicit interaction with passive affordances, IMUs and microphones have been used to detect touch events (e.g., \cite{gu2019accurate,booth2017detecting}), which can be spatially localized on the surface using multiple sensors (e.g., \cite{shi2020versatouch,gong2020acustico}).
Researchers have also examined body locations for their suitability to mount sensors, such as fingers (e.g., \cite{gu2019accurate,zhang2017fingersound}), knuckles (e.g., \cite{masson2017whichfingers,tapstrap2}), or the wrist (e.g., \cite{gong2020acustico,booth2017detecting}).

Other projects have instead researched tap detection on the user's own body.
Such projects often involve sensors arrays to detect bio-acoustic waves and are mounted on the forearm (e.g., for forearm and hand taps~\cite{harrison2010skinput}) or the wrist (e.g., for taps on finger knuckles and other body parts~\cite{chen2019taprint,yoshimura2018preliminary}).
Beyond tap events, AudioTouch uses two piezoelectric elements on the back of the hand to resolve touch force and classifies 12 hand poses during touch~\cite{kubo2019audiotouch}.
Similar to \projname{}, Actitouch complements VR interaction to detect touch input on the body~\cite{zhang2019actitouch}, using a wearable with skin contact to modulate an RF signal onto the body that a receiver picks up to detect on-body touch, while the headset supplies locations through hand tracking.

\rwsection{Finger identification during touch}
Several efforts have shown the relevance of finger identification during interaction.
On optical tabletops, such efforts used fiducial markers~\cite{marquardt2010caused} for finger-specific UI control or fingerprint scanning~\cite{holz2013fiberio} for authentication during interaction.
Gupta and Balakrishnan instead mounted optical sensors to the fingertips to distinguish two fingers, using the new capability to shrink touchscreen keyboards in half and overloading each key with two letters for finger-specific input~\cite{gupta2016dualkey}.

Body-worn IMUs are capable of finger identification when mounted close to the source, i.e., the individual fingers.
Whichfingers detected tapping fingers from piezoelectric transducers, in one embodiment placed on the fingertips and in another mounted on the knuckles~\cite{masson2017whichfingers}.
Similar in design, TapStrap is a consumer product with 5 loosely coupled rings that embed IMUs~\cite{tapstrap2}. Using them, TapStrap implements a one-handed chorded keyboard through tap detection on passive surfaces.
Booth and Goldsmith’s wrist-worn 8-sensor array along the bottom of the wrist detected gesture input~\cite{booth2018wrist}, although finger detection accuracies were below .55 across users.

Many scenarios benefit from unencumbered hands and fingers, which has brought about alternative sensing approaches.
For example, Becker et al. used a Myo band worn by the elbow and detected individual fingers from electromyography signals~\cite{becker2018touchsense}, reaching accuracies up to .75 (within-user), which slipped below .50 across users.
Also using EMG, Benko et al. classified two fingers with .91 accuracy and demonstrated use-cases on touch tables~\cite{benko2009enhancing}.

Having studied the related work in these areas, we conclude that previous approaches show the promise of finger-specific interaction.
However, they typically trade off convenience (i.e., positioning sensors close to the fingers or requiring the user to wear a glove) with accuracies that are suitable for reliably distinguishing input from five fingers.
We believe that \projname{} is an encouraging approach in this regard, mounting sensors by the user's wrist inside a convenient and socially acceptable form factor.
TapID reliably reaches high cross-user detection accuracies after minimal refinements, which adequately sets up our approach for interactive use.

\rwsection{Touch interaction in VR}

Surface touch interaction has found interest in the Mixed Reality communities.
For the purpose of text entry in VR, a commercial high-fps Optitrack system proved as a useful input platform with 23 motion-capture markers affixed to each hand~\cite{knierim2018physical}.

Without a stationary multi-camera setup in an MR scenario, the temptation has been to use the cameras built into headsets to simultaneously estimate hand poses and detect touch events.
For example, MRTouch combines the depth and infrared cameras in a Microsoft Hololens for real-time surface touch detection~\cite{xiao2018mrtouch} and discussed the main challenges in their evaluation: the high rate of missed touches (3.5\%) and spurious extra touches (19\%).

Detecting touch using cameras has been a long-standing challenge.
On his Digital Desk, Wellner explored this ambition in the early 90s~\cite{wellner1993interacting}, but concluded that reliable contact sensing required surface instrumentation.
Agarwal et al. detected touch locations and moments of contact with overhead stereo cameras~\cite{agarwal2007high}, discussing how the noise from stereo depth causes erroneous proximity readings.
Wilson later introduced an approach that built on a single depth camera, which provided more reliable detection and also scaled to mobile scenarios and on-body input (e.g., Imaginary~Phone~\cite{gustafson2011imaginary}).

These camera-based approaches share the benefit of recognizing accurate touch \emph{locations}, but all the struggle with noisy depth estimates for reliable \emph{event detection} and their constraints on input speed that is limited by camera framerates and applied filters.
This challenge is what we seek to complement with \projname{}, using wrist-worn inertial sensors that deliver the missing piece of information on rapid touch interaction: precise events and finger identification.

TapID also derives inspiration from Substitutional Reality, i.e., adopting passive affordances in the environment for VR applications~\cite{simeone2015substitutional}.
Such affordances have been used to provide otherwise missing haptic feedback in response to input events (e.g., Sparse Haptic Proxy~\cite{cheng2017sparse}).
The settings that we envision \projname{} to be used in follow our previous work on situating virtual reality interactions in a different physical reality (e.g., VRoamer~\cite{cheng2019vroamer} and Dreamwalker~\cite{yang2019dreamwalker}). 

\begin{figure}[b]
    \centering
    \includegraphics[width=\columnwidth]{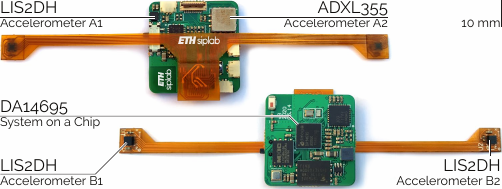}
    \caption{%
        \projname{} electronics platform.
        The main PCB features a microcontroller that reads out the two low-power IMUs that are mounted on the ends of a flexible PCB.
        For comparison in our evaluation, two additional IMUs are mounted on the main PCB.
        }
    \label{fig:electronics}
\end{figure}

\begin{figure*}[!t]
    \centering
    \includegraphics[width=\textwidth]{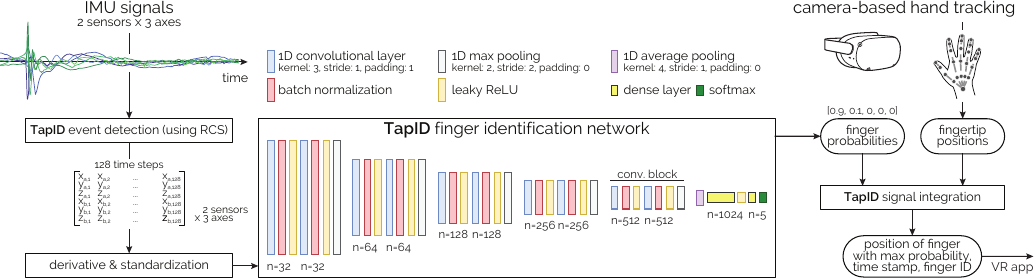}
        \caption{
            \projname's processing pipeline.
            Our primary input is the continuous stream from two 3D accelerometers, from which TapID first detects the occurrence of a tap using a rate-of-change score.
            Upon detection, we extract a window of 128 samples, centered on the tap event, and input it into our neural network classifier to identify the tapping finger.
            Our architecture comprises five blocks of convolutional layers and two final linear layers that feed into a softmax activation function, which renders a probability distribution over the fingers.
            We then combine the classification with our secondary input: continuous hand tracking.
            Having determined which finger has caused the event, we retrieve its fingertip location and forward the tap event to the VR application.
        }
    \label{fig:VGG_network}\label{fig:pipeline}
\end{figure*}

\section{\projname\; Wearable device and electronics} %

\projname\ precisely detects taps on a surface and identifies the tapping finger, made possible by the integration of our custom wristband device.
Users wear two such devices (\teaserrefA), which then operate independently per hand.
Figure~\ref{fig:electronics} shows our electronics platform, which centers around a System-on-a-Chip (DA14695, Dialog Semi) that samples all inertial sensors at a frequency of 1344\,Hz and streams the IMU data to a PC through a serial connection.
\projname\ features two low-power accelerometers (LIS2DH, STMicroelectronics) embedded into a wrist strap through a flexible PCB as well as two on-board accelerometers for testing and comparing different sensor configurations.
We cast \projname's strap using Shore-32 silicone and embedded all electronics during curing.

\section{Tap event processing pipeline} %

Our system detects and represents all tap events through a timestamp, the identity of the finger that has caused the tap, and the 2D location on the surface in world coordinates.
The system passes this record to the VR app whenever touch events occur.
In this section, we describe the three interdependent steps for determining each entry as illustrated in Figure~\ref{fig:pipeline}.
This includes our signal processing for tap detection, our machine learning-based identification pipeline, and our combination with the hand tracking from the VR headset.

\subsection{Tap event detection \& timestamp calculation}

To distinguish tap events from other arm and hand movements, we select events by the sharp spikes that they generate in the accelerometer signal.
Rather than performing na\"ive magnitude thresholding, \projname\ accumulates the rate of change in the raw signals ($RS$) of all accelerometer axes to calculate a rate-of-change score ($RCS$).
To this, we add an exponential decrease to focus on fast changes on the signal only.
\projname\ detects tap events using a threshold on $RCS$ peak prominence dependent on the last local minimum. 

\begin{equation}
\label{eq:var_score}
RCS_t = \frac{RCS_{t-1}}{1.6} + \sum_{\text{sensors}} \; \sum_{\text{axes}} \left| RS_t - RS_{t-1} \right|
\end{equation}

We derive the exact \timestamp\ of a tap event as the highest $RCS$ spike in a window of 20\,samples (15\,ms) after a threshold has been exceeded.
After each event, our processing incurs a back-off period of 200\,ms before detecting fresh events. 
This prevents a single tap from triggering multiple events while still allowing up to 600 input events per minute when using both hands.

\subsection{Tap event finger identification}

Upon detecting a the occurrence of tap event including its refined \timestamp, \projname forwards a symmetric window of raw accelerometer samples around the event's \timestamp\ to our neural network classifier to identify fingers.
Our classifier estimates the likelihood of each finger to have caused the recorded signal.
Using supervised learning, we train the network by adjusting its weights via backpropagation to minimize the cross-entropy loss on a training dataset that has a single finger label assigned to each recorded tap event.
Note that we merely train on five such classes, one for each finger, creating a classifier that is independent of individual hands.

We observed the best performance on the given task with our classifier implemented as a multi-layer feed-forward 1D-convo\-lutional neural network following a VGG-style architecture~\cite{simonyan2014very}.
Figure~\ref{fig:VGG_network} shows our convolutional architecture, which is a suitable choice given the fixed-length of the input window and the strength of convolutional layers to capture local dependencies in a signal.
To increase the receptive field, we make use of a deeper network with smaller kernels.
Intuitively, this enables the network to extract more powerful features due to the additional non-linearities between layers compared to a shallow network with larger kernels. 
Specifically, the network consists of five blocks, each made up of two convolutional layers with a kernel size of 3 that are connected to a max pooling layer to reduce the input along the time dimension by a factor of 2. 
The convolutional blocks are followed by an adaptive average pooling layer and two linear layers.
The output of the final linear layer feeds into a softmax activation function that estimates a probability distribution over the five fingers of the hand.
We represent the accelerometer data in the time domain as an input tensor of dimensions [\textit{stacked axes of sensors $\times$ window size}], where the convolution is applied over the time axis and the per-frame stacked sensor data are treated as input channels.

We apply several data pre-processing steps to boost \projname's performance. 
Taking advantage of the symmetry between the left and the right hand, we mirror the data of the left hand to receive a comparable input signal from both hands.
We swap the data of the symmetrically arranged accelerometers and invert the $x$ or $y$ axis depending on the sensor's orientation.
This reduces the variability of the combined dataset and aids the learning process of our classifier, which can now be applied to the signals from either wristband.

Next, we estimate the jerk experienced by each sensor by applying the forward difference operator along each time axis.
This operation can be thought of as finding the discrete derivative of the signal,
which amplifies high-frequency components in the Fourier domain that we argue contain the discriminative finger information we seek.
This step also reduces the lower-frequency artifacts caused by arm and hand motions.
Finally, we standardize the resulting segments along the time dimension of each channel across the dataset.

The complete classification network comprises more than 2.1 million trainable parameters.
We implemented the model in \textit{Pytorch} and trained it over 30 epochs with a batch size of 16, using the Adam optimizer~\cite{kingma2014adam} with a learning rate of $10^{-4}$, $\epsilon=10^{-8}$ and $\beta=(0.9,0.999)$.
Furthermore, we experimented with different segment sizes around the peak event and achieved the best performance with a window length of 128 frames ($\sim$95 ms).
We optimized these hyperparameters over a dataset spanning multiple users and sessions.

\subsection{Combining hand tracking with \projname\ events}

When using \projname\ in a widget-based VR user interface (e.g., those using buttons, sliders, checkboxes, etc.), we need to retrieve the location of a tap following \projname's detection of its occurrence.
To achieve this, we complement \projname's classification output with commodity hand tracking that is integrated into contemporary headsets.
Having determined which finger has caused a tap, we retrieve the fingertip position from the set of tracked hands and assign it in \projname's \textit{tap event}.
If the tip of the detected finger is not within proximity of a surface in VR, our system simply rejects detected events as inadvertent input.
In our experiments, we found 30\,mm to be a viable maximum surface distance for reliable detection and rejection.
Valid surfaces that trigger an input event in VR include the user's own body, such that our system can detect on-body touch input when UI elements are anchored to the user's arm (\teaserrefD).
In such cases, our system compares fingertip positions with the location of arm surfaces, which we approximate from the 3D-tracked joint with a constant radius (40\,mm) plus the delta region mentioned above.

Our VR system is implemented using Unity's 3D game engine for rendering and Velt~\cite{fender2018velt} for the overall data flow.
An Oculus Quest headset renders our VR applications and its SDK comes with built-in inside-out tracking for headset positions and hand poses.

While we have found the tracking data from the SDK to have sufficient accuracy for determining the user's fingertip positions in our experiments, it does not afford detecting touches on passive surfaces with any practical reliability or on-body touch input with any reliability using inside-out tracking alone.
More severely, rapid finger taps manifest themselves as shaking motions in the tracked hands, though not in position updates that allow for touch detection.
Tracking such events becomes even less suitable when fingers are partially occluded from the cameras' perspective, for example when a finger is occluded by other fingers during a down motion of the hand.
We have also found that the fingertip positions slightly lag behind the true finger positions, which is tolerable for retrieving positions on the surface, but not for detecting touch input or retrieving the exact timestamp of a touch event during interaction.

For those reasons, we merely extract fingertip positions from the hand tracking, but rely on \projname\ for detecting tap events and revealing the tapping finger.
Our system then triggers the event at the 2D position obtained from projecting the fingertip position onto the respective surface where the UI is situated.

\subsection{Pipeline latency} 

Our combination of sensing and VR-enabled hand tracking introduces several sources of latency.
The continuous hand tracking itself entails some latency, which depends on the VR device and the implemented hand tracking approach.

With regard to the latency incurred by our own processing, the \textit{total} latency from the moment of the physical tap to displaying its effect in Unity is approximately 130\,ms.
We estimated this value by manually counting frames in high-frequency video recordings.
This duration encompasses the latency caused by all hardware operation, communication, processing, and output.

We empirically determined the latency of \projname's sensing, on-board processing, and communication pipeline, which averaged out to $\sim$10\,ms.
The machine learning-based classification including our buffered data acquisition of each tap event adds 50--60\,ms. 
Thus, the latency for detecting a tap and identifying the finger comprises: 
\begin{enumerate}[itemsep=0mm,leftmargin=4.5mm]
    \item Data in sensor FIFO: 3\,ms (max: 6\,ms)
    \item SoC processing and serial transmission to host computer: 3\,ms
    \item Tap event refinement (waiting for highest amplitude): 15\,ms
    \item Buffer until the data window is complete: 47\,ms (64 samples)
    \item Neural network inference time 2\,ms (\textit{GeForce GTX 1050 Ti}) 
\end{enumerate}

Note that Steps~3 and~4 overlap by around 10\,ms on average, as the refined peak usually lies at the beginning of the refinement window.
In addition, the latency of transmitting data through serial depends on buffer settings on the host computer.

Taken together, our combination of low-latency tap input with inside-out fingertip tracking is what uniquely enables users to reliably interact with VR applications through touch, using the rapid input they know from regular touch devices.
We demonstrate several use-cases that leverage our low-latency approach in Section~\ref{sec:apps}.

\section{Technical evaluation} %
\label{sec:evaluation}

\begin{figure}[t!]
    \centering
    \includegraphics[width=\widthnarrow]{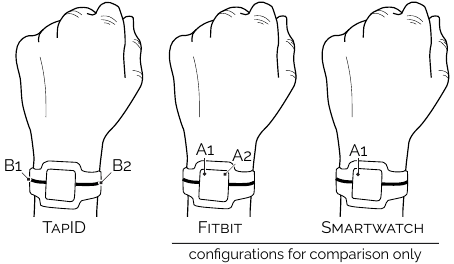}
    \caption{
        By default, \projname{} derives its signal from the IMUs $B_1,B_2$, which are embedded inside the flexible silicone strap at either side of the wrist (\tapidconfig{} configuration).
        For comparison in our evaluation, we included the two IMUs $A_1, A_2$ on the main PCB akin to \fitbitconfig{}'s Charge~\cite{fitbit_charge}.
        In the \smartwatchconfig{} configuration, only $A_1$ is active.
        Lastly, we used \allfourconfig{} IMUs as a baseline for comparison (not shown).
    }
    \label{fig:sensor_configs}
\end{figure}

\begin{figure*}[t]
    \centering
    \includegraphics[width=\textwidth]{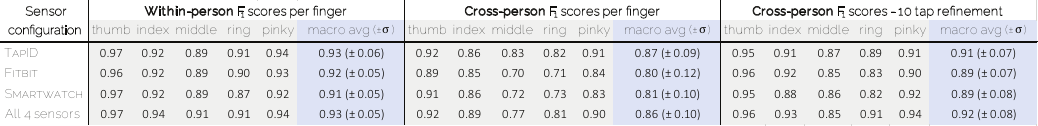}
    \caption{$F_1$ scores for \textit{cross-session} (within-person), \textit{cross-person}, and \textit{cross-person with 10-tap refinement} evaluations. Mean $F_1$ scores are broken into individual fingers and the mean and standard deviation ($\sigma$) of the \textit{Macro $F_1$ score} across participants.}
    \label{fig:F1scoreresults}
\end{figure*}

The goal of this evaluation was to verify our sensing principle and signal processing approach of detecting finger tap events and identifying the finger causing it.
We also aimed to determine to what extent wrist-mounted accelerometers in different locations and combinations contribute to the accuracy of our method. 

For this purpose, we compared the $F_1$ scores of classifying inertial data in the three configurations shown in Figure~\ref{fig:sensor_configs}: \tapidconfig{} (two accelerometers embedded in the elastic silicone strap), \fitbitconfig{} (two accelerometers rigidly mounted on the main PCB), \smartwatchconfig{} (a single low-noise board-mounted accelerometer).
We also report the $F_1$ scores for a baseline where \allfourconfig{} sensors were simultaneously used.
The second purpose of the data capture was to establish a corpus for training the cross-person tap classifier that we used for online operation in our demo apps.

We conducted the evaluation in two parts;
In Part~1, we investigated the accuracy of tap event detection.
In Part~2, we assessed the accuracies achieved by our finger identification network.
Both parts of the evaluation shared the same task and procedure, but were conducted with different participants.

\subsection{Task}

Participants comfortably sat in an office chair in a table setting, resting their lower arms on the edge of the table, and repeatedly tapped on the tabletop with the finger specified by the experimenter.
No other instructions were provided and participants controlled their own tapping frequency, intensity, and tapping locations, finger and hand motions.
The experimenter was careful to simply use the word `tap' to refer to all touch events during the experiment.

\begin{figure}[t]
    \centering
    \includegraphics[width=\widthnarrow]{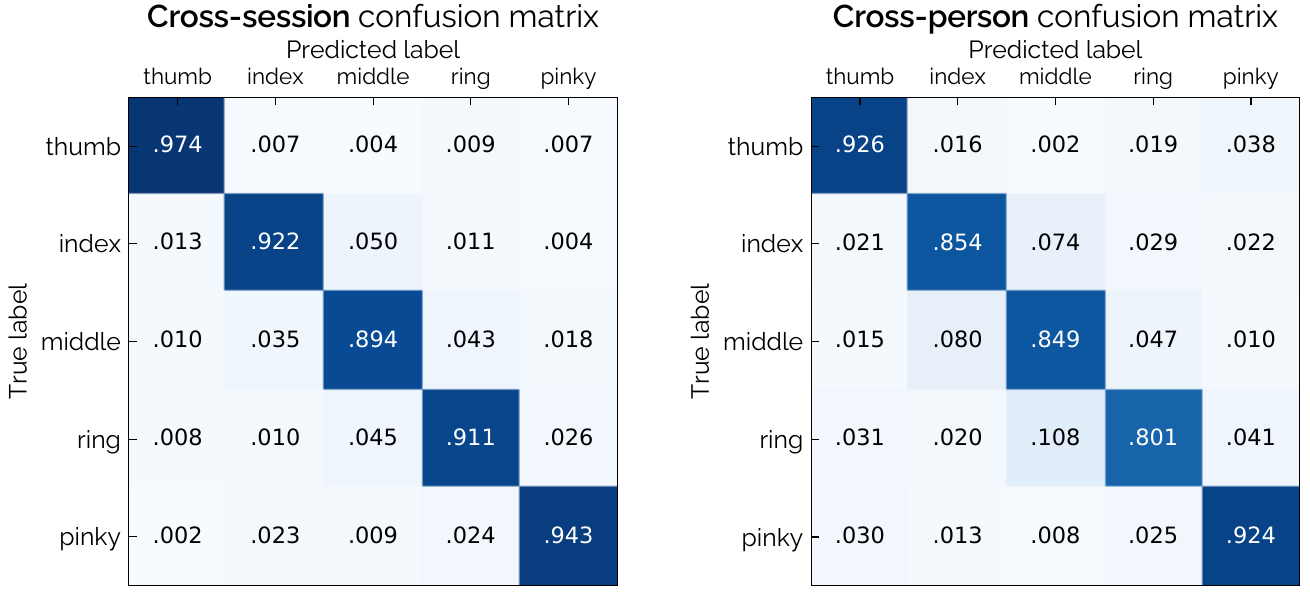}
    \caption{\textit{Left:} Normalized confusion matrix for 3-fold cross-validation for cross-session identification. \textit{Right:}~18-fold cross-validation for cross-person identification.}
    \label{fig:CMmatrices}
\end{figure}
\begin{figure}[!b]
    \centering
    \includegraphics[width=\columnwidth]{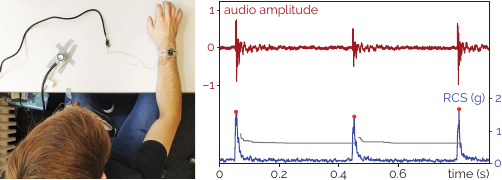}
    \caption{
        Our evaluation apparatus and procedure.
        Left: A seated participant, resting his lower arm on the table.
        The surface-mounted stethoscope served ground-truth for tap events.
        Right: The resulting audio signal and our Rate-of-Change score to detect events.
    }
    \label{fig:tapping_study}
\end{figure}

\subsection{Procedure}

The study started with a brief introduction of \projname{}, which the experimenter related to a regular smartwatch.
The experimenter then noted down the participant's age and gender and measured their wrist circumference.
Throughout the study, the experimenter instructed participants on which arm to wear the band during each session and which finger to use for tapping each round.

For each round, participants put the \projname{} device on the specified wrist and repeated 30~taps with each finger, as instructed, before taking off the band again.
A block in each session consisted of 30 taps for each finger on either hand, totaling $30 \times5~\text{fingers}\times2~\text{arms} = 300~\text{taps}$ per block.
Participants performed three such blocks with short breaks in between and produced a total of  $300~\text{taps} \times 3~\text{blocks} = 900$ taps. 
Each participant completed the study in under 20 minutes.

\subsection{Participants}

We recruited 18 participants (14 male, 4 female, ages 19--57, mean = 28.4 years). Participants' average wrist circumference was 168\,mm (SD = 10\,mm, min = 146\,mm, max = 185\,mm). Each participant received a small gratuity for their time.
Of the 18 participants, 4 (1 female, ages 23--29, mean = 25.3 years) additionally took part in evaluating the performance of our tap detection. Their wrist circumferences ranged from 158\,mm to 176\,mm.

\subsection{Evaluating tap event detection}

To record ground truth annotations for surface taps, we recorded a surface acoustic signal using a stethoscope that was covered and taped to the tabletop (Figure~\ref{fig:tapping_study}). 
During the evaluation, participants kept quiet to avoid introducing artifacts into the recorded signal.

This sub-evaluation resulted in a total of 3600 tap events. 
Using our RC score as a measure (Equation~\ref{eq:var_score}), our tap detection achieved an accuracy of \fone{tap}{.996} (precision 99.7\%, recall 99.5\%) in the \tapidconfig{} configuration.
For \fitbitconfig{} and \smartwatchconfig{}, the detection achieved \fone{tap}{.996} as well.

\begin{figure}[t]
    \centering
    \includegraphics[width=\widthnarrow]{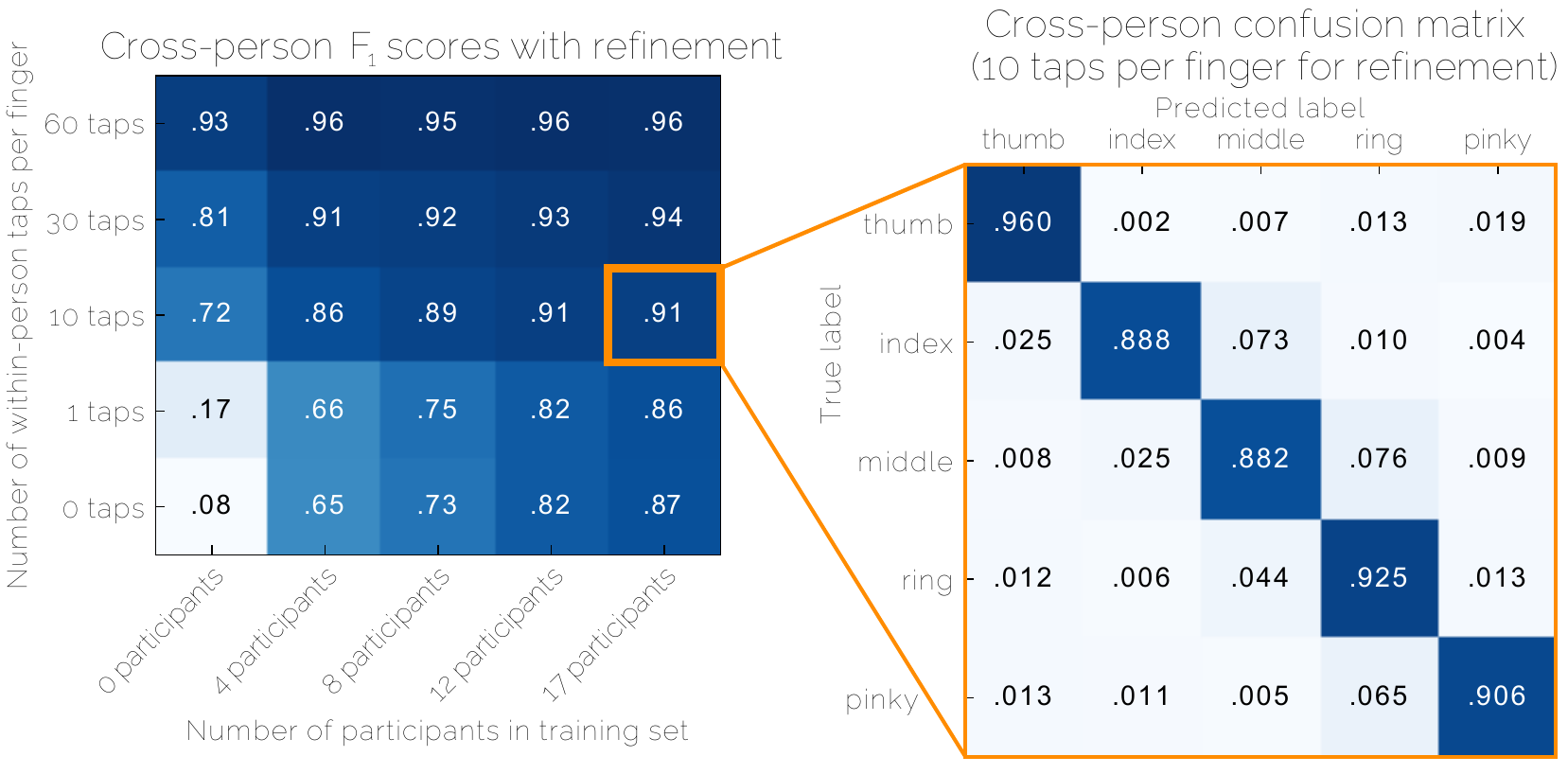}
    \caption{Cross-person refinement accuracy. \textit{Left:}~$F_{1}$ scores depending on training and refinement set sizes. \textit{Right:}~Normalized confusion matrix with 18-fold cross-validation with 10 within-person taps per finger for refinement.}
    \label{fig:refinementexperiment}
\end{figure}

\subsection{Evaluating tap event identification}

In this sub-evaluation, ground-truth resulted from the experimenter's instructions which finger to use for tapping, which the experimenter logged in our data collection tool for post-hoc label assignment.

\subsubsection{Results in \tapidconfig{} configuration}

We calculate the average $F_1$ score for each finger across all participants.
Moreover, we report the mean and the standard deviation of the \textit{Macro $F_1$} score across all participants, which treats all fingers as being equally important by taking the unweighted mean across class-wise $F_1$ scores.

\textit{Cross-session~identification~accuracy:}
We split the recorded data by participant and block.
For each participant, we performed 3-fold cross-validation, using two blocks for training and one block for testing while ignoring all data from other participants.
Across all participants, the accuracy was, on average, \fone{finger}{.93} (SD=0.06).

\textit{Cross-person~identification~accuracy:}
We performed 18-fold cross-validation, testing on each participant's events and training on all other participants.
The accuracy was \fone{finger}{.87} (SD=0.09).

\textit{Cross-person~identification~accuracy~with~refinement:} 
In addition to pure cross-person testing, we assessed the impact of refinement with individual tap events from a participant's set on the cross-person identification accuracy $F_{1,\text{finger}}$.
Simultaneously, we evaluated the impact of the size of the training dataset on the average cross-person identification accuracy.

\autoref{fig:refinementexperiment} shows the mean $F_{1}$ scores across participants depending on number of random taps used for refinement (rows) and depending on number of participants' data used for training (columns).
For this evaluation, we first trained the classifier for 30 epochs on the recorded taps of (i)~0, (ii)~4, (iii)~8, (iv)~12 and (v)~17 randomly drawn participants---excluding the person in the test set.
We then fine-tuned the network for another 30 epochs using only the first (a)~1, (b)~10, (c)~30, and (d)~60 tap events per finger of the tested participant and removed these samples from the test set.
We tested the refined classifier on the remaining blocks for that participant.

After fine-tuning, our trained network produced $F_{1}$ scores that rose with the number of participants in the training set as well as with the number of refinement taps.
With 18-fold cross-validation and training with (iv)~all 17 participants' training data, our classifier achieves a mean $F_{1}$ score of (a)~\fone{finger}{.86} (SD=0.07), (b)~\fone{finger}{.91} (SD=0.07), (c)~\fone{finger}{.94} (SD=0.05), and (d)~\fone{finger}{.96} (SD=0.05) after corresponding refinement.

\subsubsection{Results in other configurations} %

For our baseline comparisons, Figure~\ref{fig:F1scoreresults} lists all the scores achieved with the corresponding sensor configurations alongside a baseline using \allfourconfig{} IMUs. %
Compared to the \fitbitconfig{} and \smartwatchconfig{} configurations, \tapidconfig{} produced consistently higher $F_1$ scores.
In comparison, using \allfourconfig{} IMUs for training and testing added marginal improvements over \tapidconfig{} and only in some of the cases.

\subsection{Discussion}

The evaluation confirmed the validity of our sensing principle and our implementation on a wrist-worn device to capture body-coupled events.
It also confirmed the \tapidconfig{} configuration of our prototype, embedding two low-cost, low-power IMUs in the silicone strap.

Regarding the tap detection, the accuracy \projname{} achieved is high enough for practical purposes, but it also is in \fitbitconfig{} and \smartwatchconfig{} configuration.
This confirms recent results in related efforts (e.g.,~\cite{gu2019accurate}), though with IMU sensors positioned at the wrist for unencumbered hands instead of placing them on the fingers.
One trade-off remains in our tap detection: Decreasing the back-off period allows for faster interactions but increases the likelihood that one (strong) tap produces two events (i.e., one false positive).

As for the performance of our classifier to identify individual fingers, unsurprisingly we saw the best cross-person accuracy when training with the largest amount possible, refined with person-specific training data, reaching $F_1$ scores of up to .96 this way (Figure~\ref{fig:refinementexperiment}).
What is notable, however, is that a single person-specific calibration procedure holds across recording sessions, setting up novel users with a one-time on-boarding step to reach an accuracy of $.96$.
Even by just fine-tuning with 10 samples, which can easily be recorded in under a minute, \projname{} identifies the correct finger with an accuracy of \fone{finger}{.91}, notably for any given touch event.

Without refinement, however, cross-person classification achieves an accuracy of \fone{finger}{.87}, which we believe still has room for improvement to rise to practical levels.
That said, this $F_1$ score compares favorably to results achieved in the related work with comparable setups with unencumbered fingers and hands (e.g.,~\cite{becker2018touchsense, booth2018wrist}) for cross-person classification.
Interestingly, our method achieves the largest performance increase in \tapidconfig{} configuration over the \fitbitconfig{} and \smartwatchconfig{} configurations (Figure~\ref{fig:F1scoreresults}), while person-specific refinement narrows the gap between configurations.

Taking a look at the performance of individual fingers, our method most reliably identifies thumb and pinky taps, while distinguishing between middle and ring fingers proved to be more challenging.
While all finger $F_1$ scores are $>.89$ for cross-session, the ring finger drops to $.82$ for cross-person $F_1$ scores.
The right confusion matrix in \autoref{fig:CMmatrices} illustrates that the model especially struggles to differentiate between the index, middle and ring finger.

Refinement with just 10 taps, however, leads to an improved discrimination between the three inner fingers of the hand and recovers all finger's $F_1$ scores to values $>.87$ (see Figure~\ref{fig:refinementexperiment} and Figure~\ref{fig:F1scoreresults}).

Inspecting results in detail, physical measures such as a person's wrist circumference may impact the classification accuracy.
Performing cross-validation on the 9 participants with a wrist circumference closest to the average of our population, the cross-person classification without refinement rises to \fone{finger}{.92} for \tapidconfig{} (compared to \fone{finger}{.82} when using the other half of our population for validation).
Similar trends occurred for all sensor configurations
and one reason may be the constant distance between IMUs.
Based on these results, we hypothesize that training data from more participants with smaller and larger wrists will decrease this effect but strengthen our cross-person accuracy without refinement.

\newcommand\appemph[1]{\textit{#1}}
\newcommand\appfigref[1]{%
\ (Figure~\ref{fig:applications}#1)%
}%
\newcommand\appsection[1]{\textit{\textbf{#1:}}}

\FIGApps{t}

\section{Use cases}
\label{sec:apps}

\projname\ can be used for complementary high-fidelity input in many VR scenarios.
In this section, we first describe broader scenarios and envisioned use cases for \projname.
We then showcase several apps that we implemented to demonstrate the usefulness of our approach.

\subsection{Scenarios} %

In typical VR scenarios, users interact with non-tangible objects, which lack all physical affordances and, thus, require interaction that is mediated through hand-held controllers or mere hands.
However, mid-air input---especially in the case of prolonged interaction---often triggers fatigue, because the lack of physical affordances leaves no options for rest~\cite{simeone2015substitutional}.
We argue that, therefore, the main success of VR applications so far has been in gaming that engages the whole-body (e.g., Beat Saber~\cite{beatsaber}).
In contrast, even the most recent productivity applications that attempt to replicate office environments (e.g., Facebook Infinite Spaces~\cite{infiniteoffice}) focus on content \textit{consumption} rather than content creation.
We believe this is a direct result of the inadequate interaction modality that is mid-air input, often retrofit to an interface that was designed for keyboards, mice, and not least touch.
In VR, no devices currently exist to facilitate such interaction other than the handheld controllers, which induce short-term novelty, but have so far not been convincing in enabling long-term productivity, especially when compared to conventional desktop input devices.

We believe that the potential of immersive VR productivity apps remains untapped if the input modality is not effective, precise, and non-fatiguing.
Much of productivity work is characterized by sustained interaction, often in the form of fluid and rapid input, often bimanual, and often involving tools to accomplish tasks.
Our prototype \projname\ now offers an alternative interaction concept, which is inspired by interaction in physical offices, but adapted to support potentially infinite virtual spaces of content in VR.

Besides productivity-based scenarios, \projname\ can also be useful in more playful use-cases that differ from previous (mostly mid-air based) VR experiences.
For instance, \projname\ enables VR games that require surface input with exact timing (e.g., rhythm games).

\subsection{Application prototypes}

We implemented several applications that showcase our unique combination of rapid tap interaction in VR through \projname\ in conjunction with spatial interfaces mediated by a VR headset.
In these demonstrations, the user is situated in VR and surrounded by content typical in desktop and touch user interfaces, such as photo collections and web browsers.
To facilitate interaction with the user interface, we developed a series of \textit{tapping-based virtual input devices} that the user can open and position by tapping the desired location on the surface with specific fingers.
For instance, the left pinky finger can summon arrow keys while the right pinky finger triggers a numeric keypad to appear and the thumb can summon a simple keyboard between both hands.

While our applications show hands during interaction, we render fingertips and touches using colored spheres during ego-centric use, showing a line that projects down to the surface and following previous UI recommendations~\cite{grubert2018effects}.
Screenshots of our applications are shown in Figures~\ref{fig:teaser} and~\ref{fig:applications}.

\appsection{\appphoto}
Familiar from multi-touch devices, we implemented a \appemph{\appphoto} app as shown in Figure~\ref{fig:applications}a).
The app uses \projname's reliable tap detection to trigger drag events for moving and resizing pictures.
A command input menu enables step-wise resizing and rotating images.
With our approach, we can easily assign these commands to each finger.
This prototype also works without a table in an on-body configuration with one wristband worn on one arm and with the user's other arm as tapping surface (see \teaserrefD).

\appsection{\appdocedit}
We designed a simple keyboard input widget to add text to documents (Figure~\ref{fig:applications}b), complementing the use of existing document authoring tools (e.g., Microsoft Word, PowerPoint).
Words can be selected through double-tap, e.g., to change their format.
Additionally, our arrow key virtual input (summoned with left pinky finger) aids in quickly adjusting the text cursor's desired location.
This demonstrates our support of conventional touch capabilities, seamlessly integrating our finger-specific tapping-based virtual input devices with operating system-provided controls and existing applications that were designed for touch.

\appsection{\appnumberinput}
Users can open a virtual numeric keypad widget to type in numbers and operations as shown in Figure~\ref{fig:applications}c.
This widget conveniently fits below the hand, so that users can quickly input numbers using all five fingers of one hand.
This affordance is useful, for example, when summoning and using a calculator as part of the interaction flow in a user interface.

\appsection{\apppiano}
Our final demonstration is a virtual piano app as shown in Figure~\ref{fig:applications}d.
This VR piano optimally leverages \projname\ for \emph{temporally precise input} and enables users to hit piano keys using any finger.
The combination of visual feedback in VR, the passive haptic feedback of the surface, and the low latency of our approach uniquely enables playing this virtual instrument while preserving the exact beat of the user's tapping input.

\section{Limitations}

Before reaching production-level performance, a couple of limitations remain in our current implementation.
Apart from the currently necessary user-specific refinements to achieve reliable accuracies, we have not tuned or evaluated our method on its capability to identify multi-touch events---at this time, only individual fingers are detected, though with simultaneous bimanual input.
A second assumption behind our current implementation is the use of a table or the user's own body as a passive input surface.
We did not experimentally and quantitatively evaluate the detection accuracy of our current implementation on surfaces other than office tables made from coated wood.
We also did not explicitly vary or control tapping strength.
We have tested and evaluated our system in a stationary and seated setting, without accounting for body motion during use (e.g., in mobile scenarios).
With respect to supported input poses, \projname{} detects and identifies taps with bent fingers.
In this case, however, the derived input location may be inaccurate or even unavailable and thus unknown due to our current reliance on the input locations reported from the headset.
For our current system to work, the hand producing the input event must be in the tracking area of the VR headset, otherwise input events on the wrist-worn sensor will be dismissed as inadvertent.
This entails that, currently, the user needs to look at the content they interact with in VR, including text input on the shown keyboard, which, thus, does not support eyes-free operation.
All interaction is also required to be direct, which previous work has found to lead to higher error in 3D scenes~\cite{simeone2015comparing}, but work by Knierim et al. on text input in VR showed to be beneficial, as experienced typists improved their typing performance by seeing their hands and the keyboard in VR~\cite{knierim2018physical}.

Regarding the suitability of our approach to be implemented on commodity smartwatches, we acknowledge that it may appear unreasonable for a user to wear two smartwatches, which our current approach requires for bimanual operation.
However, much like currently used controllers for input, it may seem fathomable that future users put on two small wristbands for all interaction in VR.
Finally, our current inference pipeline relies on a moderate GPU and is, at this point, not suitable for embedded processing.

\section{Conclusions}

We have presented \projname{}, a wearable sensing device that complements camera-based hand tracking to bring rapid touch interaction to VR systems.
Our approach, thus, establish the basis for reliable and prolonged interaction during continuous use, where users may rest their arms and hands on a surface.
For each tap on a passive surface, such as a tabletop, \projname{} reliably detects the event and identifies which finger has been used for the touch.
This allows VR systems that track the user's hands in full to add the missing key element: rapid surface contact detection that additionally disambiguates which finger caused the action, such that the VR app can trigger the input event in the application.

In our evaluation with 18 participants, \projname{} produced convincing accuracies for tap detection and finger identification in a cross-session setting.
Cross-validation testing with a classifier that was trained on just cross-person data showed a drop in accuracy, from which \projname can recover, however, with a small number of a user's tap events for refinement training.

Overall, we believe that the approach we have outlined in this paper is a viable complement for existing VR systems that allows the familiar input modality that is touch input to transfer to spatial VR systems, potentially even during everyday scenarios.
A promising insight in this regard is the evaluation of our method in \fitbitconfig{} configuration, using only \projname's board-mounted IMU sensors as found in existing commodity wearables~\cite{fitbit_charge}, which produced comparable accuracies when implementing our method.

\bibliographystyle{SIGCHI-Reference-Format}
\bibliography{Tappo-base}

\balance

\end{document}